\documentstyle[twoside,fleqn,espcrc2,epsfig]{article}
\newread\epsffilein 
\newif\ifepsffileok 
\newif\ifepsfbbfound 
\newif\ifepsfverbose 
\newdimen\epsfxsize 
\newdimen\epsfysize 
\newdimen\epsftsize 
\newdimen\epsfrsize 
\newdimen\epsftmp  
\newdimen\pspoints  
\def\epsfbox#1{\global\def\epsfllx{72}\global\def\epsflly{72}%
 \global\def\epsfurx{540}\global\def\epsfury{720}%
 \def\lbracket{[}\def\testit{#1}\ifx\testit\lbracket
 \let\next=\epsfgetlitbb\else\let\next=\epsfnormal\fi\next{#1}}%
\def\epsfgetlitbb#1#2 #3 #4 #5]#6{\epsfgrab #2 #3 #4 #5 .\\%
 \epsfsetgraph{#6}}%
\def\epsfnormal#1{\epsfgetbb{#1}\epsfsetgraph{#1}}%
\def\epsfgetbb#1{%
\openin\epsffilein=#1
\ifeof\epsffilein\errmessage{I couldn't open #1, will ignore it}\else
 {\epsffileoktrue \chardef\other=12
 \def\do##1{\catcode`##1=\other}\dospecials \catcode`\ =10
 \loop
  \read\epsffilein to \epsffileline
  \ifeof\epsffilein\epsffileokfalse\else
   \expandafter\epsfaux\epsffileline:. \\%
  \fi
 \ifepsffileok\repeat
 \ifepsfbbfound\else
 \ifepsfverbose\message{No bounding box comment in #1; using defaults}\fi\fi
 }\closein\epsffilein\fi}%
\def\epsfclipstring{}
\def\epsfsetgraph#1{%
 \epsfrsize=\epsfury\pspoints
 \advance\epsfrsize by-\epsflly\pspoints
 \epsftsize=\epsfurx\pspoints
 \advance\epsftsize by-\epsfllx\pspoints
 \epsfxsize\epsfsize\epsftsize\epsfrsize
 \ifnum\epsfxsize=0 \ifnum\epsfysize=0
  \epsfxsize=\epsftsize \epsfysize=\epsfrsize
  \epsfrsize=0pt
arithmetic!
  \else\epsftmp=\epsftsize \divide\epsftmp\epsfrsize
  \epsfxsize=\epsfysize \multiply\epsfxsize\epsftmp
  \multiply\epsftmp\epsfrsize \advance\epsftsize-\epsftmp
  \epsftmp=\epsfysize
  \loop \advance\epsftsize\epsftsize \divide\epsftmp 2
  \ifnum\epsftmp>0
   \ifnum\epsftsize<\epsfrsize\else
    \advance\epsftsize-\epsfrsize \advance\epsfxsize\epsftmp \fi
  \repeat
  \epsfrsize=0pt
  \fi
 \else \ifnum\epsfysize=0
  \epsftmp=\epsfrsize \divide\epsftmp\epsftsize
  \epsfysize=\epsfxsize \multiply\epsfysize\epsftmp   
  \multiply\epsftmp\epsftsize \advance\epsfrsize-\epsftmp
  \epsftmp=\epsfxsize
  \loop \advance\epsfrsize\epsfrsize \divide\epsftmp 2
  \ifnum\epsftmp>0
  \ifnum\epsfrsize<\epsftsize\else
   \advance\epsfrsize-\epsftsize \advance\epsfysize\epsftmp \fi
  \repeat
  \epsfrsize=0pt
 \else
  \epsfrsize=\epsfysize
 \fi
 \fi
 \ifepsfverbose\message{#1: width=\the\epsfxsize, 
height=\the\epsfysize}\fi
 \epsftmp=10\epsfxsize \divide\epsftmp\pspoints
 \vbox to\epsfysize{\vfil\hbox to\epsfxsize{%
  \ifnum\epsfrsize=0\relax
  \includegraphics{#1}%
  \else
  \epsfrsize=10\epsfysize \divide\epsfrsize\pspoints  
  \includegraphics{#1}%
  \fi
  \hfil}}%
\global\epsfxsize=0pt\global\epsfysize=0pt}%
 {\catcode`\%=12
\global\let\epsfpercent=
\long\def\epsfaux#1#2:#3\\{\ifx#1\epsfpercent
 \def\testit{#2}\ifx\testit\epsfbblit
  \epsfgrab #3 . . . \\%
  \epsffileokfalse
  \global\epsfbbfoundtrue
 \fi\else\ifx#1\par\else\epsffileokfalse\fi\fi}%
\def\epsfempty{}%
\def\epsfgrab #1 #2 #3 #4 #5\\{%
\global\def\epsfllx{#1}\ifx\epsfllx\epsfempty
  \epsfgrab #2 #3 #4 #5 .\\\else
 \global\def\epsflly{#2}%
 \global\def\epsfurx{#3}\global\def\epsfury{#4}\fi}%
\def\epsfsize#1#2{\epsfxsize}
\let\epsffile=\epsfbox
\newcommand{\beq}{\begin{equation}}
\newcommand{\eeq}{\end{equation}}
\newcommand{\absvcb}{\vert V_{cb}\vert}

\newcommand{\absvtd}{\vert V_{td}\vert}
\newcommand{\absvts}{\vert V_{ts}\vert}
\newcommand{\absvtb}{\vert V_{tb}\vert}
\newcommand{\abseps}{\vert\epsilon\vert}
\newcommand{\epsp}{\epsilon^\prime/\epsilon}

\newcommand{\BGAMAXS}{B \rightarrow X _{s} + \gamma}
\newcommand{\BGAMAXD}{B \rightarrow X _{d} + \gamma}
\newcommand{\BBGAMAXS}{{\cal B}(B \rightarrow  X _{s} + \gamma)}
\newcommand{\BBGAMAXD}{{\cal B}(B \rightarrow  X _{d} + \gamma)}

\def\mt{m_t}
\def\mb{m_b}
\def\mc{m_c}

\newcommand{\delmd}{\Delta M_d}
\newcommand{\delms}{\Delta M_s}

\newcommand{\kkbar}{$K^0$-${\overline{K^0}}$}
\newcommand{\bdbdbar}{$B_d^0$-${\overline{B_d^0}}$}
\newcommand{\bsbsbar}{$B_s^0$-${\overline{B_s^0}}$}
\newcommand{\as}{\mbox{$\alpha_{\displaystyle s}$}}

\def\bea{\begin{eqnarray}}
\def\eea{\end{eqnarray}}
\def\be{\begin{equation}}
\def\ee{\end{equation}}

\def\g{\gamma}

\newcommand{\bgamaxs}{$B \to X _{s} + \gamma$}

\def\Vcbabs{\vert V_{cb} \vert}

\def\Vubabs{\vert V_{ub}\vert}

\def\Vtdabs{\vert V_{td} \vert} 
\def\Vtsabs{\vert V_{ts} \vert}

\newcommand{\go}[1]{\gamma^{#1}}
\newcommand{\gu}[1]{\gamma_{#1}}

\def\qbar{\overline q}

\def\q5q{\qbar{{\lambda_a}\over 2} i\gamma_5 q}

\def\ra{\rightarrow}

\newcommand{\bgamaxd}{$B \to X _{d} + \gamma$}

\def\bsll{$b \rightarrow s \ell^+ \ell^- $ }
\def\bxsll{$B \rightarrow X_s \ell^+ \ell^- $ }
\def\bxsee{B \rightarrow X_s e^+ e^-  }
\def\bxsmm{B \rightarrow X_s \mu^+ \mu^-  }
\def\bxstt{B \rightarrow X_s \tau^+ \tau^- }

\def\bxsg{$B \rightarrow X_s \gamma $ }
\newcommand{\BGAMAKSTAR}{B \ra  K^{\star} + \gamma}

\newcommand{\GGAMAXD}{\Gamma(B \ra  X _{d} + \gamma)}

\newcommand{\GGAMAXS}{\Gamma (B \ra  X _{s} + \gamma)}
\def\bxsll{$B \rightarrow X_s \ell^+ \ell^- $ }
\def\bxsee{B \rightarrow X_s e^+ e^-  }
\def\bxsmm{B \rightarrow X_s \mu^+ \mu^-  }
\def\to{\rightarrow}

\def\mb{m_b}

\def\as{\alpha _s}

\newcommand{\AmS}{{\protect\the\textfont2
  A\kern-.1667em\lower.5ex\hbox{M}\kern-.125emS}}

\begin{document}
\begin{flushright}
DESY 97-019\\
February 1997\\
\end{flushright}
\begin{center}
{\Large \bf
\centerline{Flavour Changing Neutral Current Processes in $B$ Decays}}
\vspace*{1.5cm}
 {\large A.~Ali}
\vskip0.2cm
 Deutsches Elektronen Synchrotron DESY, Hamburg \\
Notkestra\ss e 85, D-22603 Hamburg, FRG\\
 
\vspace*{8.0cm}
To be published in the Proceedings of the Fourth KEK
Topical Conference on\\
Flavour Physics, KEK, Tsukuba, Japan, October 29-31, 1996

\end{center}
\thispagestyle{empty}
\newpage
\setcounter{page}{1}
\title{Flavour Changing Neutral Current Processes in $B$ Decays}

\author{A.~Ali\address{Deutsches Elektronen-Synchrotron DESY\\ 
        Notkestra\ss e 85, D-22603 Hamburg, FRG}}
       
\begin{abstract}
We review the theory and experimental measurements in the flavour 
changing neutral current (FCNC) processes involving $B$ decays in the 
context of the standard model (SM). The role of these processes in 
determining the weak mixing matrix elements is stressed.

\end{abstract}

\maketitle

\section{The interest in FCNC Processes}

 In SM, there are no FCNC processes present at the tree level due
to the built-in GIM Mechanism \cite{GIM}.
They are induced by higher orders (Penguins and Boxes) and QCD plays an
essential role in determining their amplitudes. Their rates 
show sensitivity to higher scales. For example, the $K_L$-$K_S$ mass 
difference $\Delta M_K$
is sensitive to the charmed quark mass. The estimate of $m_c=O(2)$ GeV 
based on the analysis of $\Delta M_K$ by Ben Lee and Mary Gaillard in 1974 
was a good long 
shot \cite{LG74}. In contrast, FCNC $B$ decays are governed by the 
intermediate 
top quark state. The first indication of the large (in excess of 100 GeV)
top quark mass came from the measurement of a sizable \bdbdbar ~mixing 
by the ARGUS collaboration almost a decade ago \cite{ARGUS87}. This has 
been confirmed subsequently by the direct production experiments at the 
Tevatron \cite{top95}. With the present value $m_t=175 \pm 6$ GeV
and the projected precision on $\mt$ at 
the upgraded Tevatron $\delta m_t \simeq 1$ GeV  \cite{CDFvtb}, FCNC 
$B$ decays are probably not competitive
in the precise determination of the top quark mass. However, they do
 provide quantitative
information on the Cabibbo-Kobayashi-Maskawa (CKM)
matrix elements \cite{CKM}  
$\absvtd, ~\absvts$, and $\absvtb$.
These quantities are  hard to get in direct production and decays of 
the top quark, with the
exception of $\absvtb$ for which first measurements from CDF 
already exist $\absvtb = 0.97 \pm 0.15 \pm 0.07$ \cite{CDFvtb}, and which can
be vastly improved at the linear colliders (a precision of
$\delta \absvtb/\absvtb= 1\%$ is projected \cite{Fujiikek}).
 Hence, FCNC processes will remain in the foreseeable future the only 
means to measure the 
matrix elements $\absvtd$ and $\absvts$. Thus, 
 the interest in FCNC decays can be formulated as follows:
\begin{itemize}
\item They measure the parameters of the CKM matrix and play a central role
in testing the unitarity of this matrix.
\item They are vitally important in measuring CP violation in flavour 
changing processes. The only example (so far) of CP violation is in the
\kkbar ~sector; the quantity $\abseps$ is related to the phase in the
non-diagonal elements of the \kkbar ~mass matrix. Measurements of 
the CP-violating quantity $\epsp$ in the kaon sector and the phases 
$\alpha, ~\beta, ~\gamma$ in the $B$ sector involve 
particle-antiparticle mixings.
\item They offer fertile testing grounds for QCD; 
the available techniques (Perturbative QCD, Heavy quark effective theory 
HQET, Lattice-QCD, QCD-sumrules) are directly applicable in these decays.
\item  They may reveal new physics, such as supersymmetry SUSY,
more generations, leptoquarks, etc.
\item Most importantly, they are accessible at the present and 
planned experimental facilities 
(CESR, LEP/SLC, Tevatron, B Factories, HERA-B/LHC, K factories).
\end{itemize}
\section{Examples of FCNC processes}
FCNC $K$ decays have had a glorious past!
They had a great impact on the development of the
SM. In particular, the six-quark model of Kobayashi and Maskawa, with a 
complex phase in the quark mixing matrix, was proposed to accommodate 
$\abseps$ in the SM. The future interest in $K$ decays 
is being reviewed by Buchalla at this conference \cite{Buchallakek}.
 
One does not anticipate measurable rates involving FCNC decays of
the $D$ and $D_s$ mesons or the $\Lambda_c$ baryons in the SM. This is 
dictated by the GIM construction and the known pattern of the quark masses 
and mixing matrix.
No drastic enhancements of FCNC decays in the charmed sector are expected
in the minimal extensions of the SM either, such as supersymmetry. 
 However, in more drastic scenarios beyond the SM - such as the existence of
around the corner leptoquarks - measurable effects may show themselves in 
the FCNC charmed hadron decays. These processes include
$D^0$ - $\overline{D^0}$ mixing and CP Violation; 
$D^0 \to \mu^+ \mu^-, ~D^\pm \to \pi ^\pm \mu^+ \mu^-$ etc. Their 
experimental searches must be set forth, as emphasized at this meeting by
Golowich \cite{Golowichkek}.

Several FCNC transitions have been measured in $B$ 
decays. Many more are expected to be measured, with the 
SM-based decay rates being close to their respective experimental
upper limits.
Some of the experimentally interesting FCNC processes in $B$ 
decays are listed below.\\
$\bullet$ \bdbdbar ~and \bsbsbar ~mixings,\\
$\bullet$ Electromagnetic penguins: $\BGAMAXS$, \\ $B \to K^* + \gamma,
~\BGAMAXD,~B \to (\rho, \omega) + \gamma$\\
$\bullet$  FCNC Semileptonic Decays: $B \to X_s +\ell^+ \ell^-, ~B \to 
(K,K^*) + \ell^+ \ell^-, ~B \to X_d + \ell^+ \ell^-, ~B \to
(\pi,\rho,\omega) + \ell^+ \ell^-, ~B \to (X_s,X_d)+ \nu \bar{\nu}, ~B
\to (K,K^*,\pi,\rho) + \nu \bar{\nu}$\\
$\bullet$ FCNC Leptonic and radiative Decays:\\ $B_s^0 \to \ell^+ \ell^-,
~B_s^0 \to \gamma \gamma$.

This article briefly reviews the theory of 
several decay modes listed above. Comparisons of the SM estimates with 
experiment are made wherever such information is available. This is
used to determine the CKM matrix elements.
The possibility of observing beyond-the-SM physics
in FCNC $B$ decays is also briefly discussed.
 \section{Estimates of $\BBGAMAXS$ and $\Vtsabs$ in the Standard Model}
\par
 We start with the decay $\BGAMAXS$, which has been measured by CLEO
\cite{CLEOrare2}. This was preceded by the measurement of the exclusive 
decay $\BGAMAKSTAR$ by the same collaboration \cite{CLEOrare1}. The present 
measurements give \cite{CLEOwarsaw}:
\begin{eqnarray}
\label{penguinexp}
{\cal B}(\BGAMAXS) &=& (2.32\pm 0.57\pm 0.35)\times 10^{-4}, \nonumber\\
{\cal B}(\BGAMAKSTAR) &=& (4.2\pm 0.8 \pm 0.6)\times 10^{-5},
\end{eqnarray}
yielding an exclusive-to-inclusive ratio:
\begin{equation}
R_{K^*} = \frac{\Gamma(\BGAMAKSTAR)}{\Gamma(\BGAMAXS)}=(18.1\pm 6.8)\% ~.
\end{equation}
The decay rates in eqs.~(\ref{penguinexp}) determine the ratio of the CKM 
matrix 
elements $\Vtsabs/\Vcbabs$ and the quantity $R_{K^*}$ provides information
on the decay form factor in $\BGAMAKSTAR$. In what follows we take up
these points.

The leading contribution to $b \to s +\gamma$ arises
at one-loop from the so-called penguin diagrams. With the help of the
unitarity of the CKM matrix,
the decay matrix element in the lowest order can be written as:
\bea
\label{e2}
\lefteqn{{\cal M }(b \to s ~+\gamma)
    = \frac{G_F}{\sqrt{2}}\,\frac{e}{2 \pi^2} \,\lambda_{t}
(F_2 (x_t)-F_2(x_c))}\, 
\nonumber\\&&{}
\times  q^\mu \epsilon^\nu \bar{s} \sigma_{\mu \nu} (m_bR ~+ ~m_sL)b ~. 
\eea
where $G_F$ is the Fermi coupling constant, $e=\sqrt{4 \pi 
\alpha_{\mbox{em}}}$, $x_i= ~m_i^2/m_W^2$; 
$q_\mu$  and $\epsilon_\mu$ are, respectively, the photon four-momentum
and polarization vector.
The GIM mechanism \cite{GIM} is manifest in this amplitude and the
CKM-matrix element dependence is factorized in $\lambda_t\equiv V_{tb} 
V_{ts}^*$.
The (modified) Inami-Lim function $F_2(x_i)$ derived from the (1-loop) 
penguin diagrams is given by \cite{InamiLim}:
\bea
\lefteqn{F_{2}(x) = 
 \frac{x}{24 (x-1)^{4}}}
\nonumber\\&&{}  
\times  \left[6 x (3 x -2 )
\log x - (x-1) (8 x^{2} +5 x -7 ) \right]. \nonumber \\
\eea
 The measurement of the branching ratio for $\BGAMAXS$ can be 
readily interpreted in terms of the CKM-matrix element product
$\lambda_t/\Vcbabs$ or equivalently $\Vtsabs/\Vcbabs$.
For a quantitative determination, however,  QCD radiative
corrections have to be computed and
the contribution  of the so-called long-distance (LD) effects estimated.

  The appropriate framework to incorporate
QCD corrections is that of an effective theory obtained by integrating 
out the
heavy degrees of freedom, which in the present context are the top quark 
and $W^\pm$ bosons.
 The effective Hamiltonian depends on the underlying theory and 
for the SM one has (keeping operators up to dimension 6), 
\begin{equation}\label{heffbsg}
{\cal H}_{eff}(b \to s +\gamma) = - \frac{4 G_F}{\sqrt{2}} V_{ts}^* V_{tb}
        \sum_{i=1}^{8} C_i (\mu) {\cal O}_i (\mu) ,
\end{equation}
where the operator basis, the lowest order ``matching conditions" 
$C_{i}(m_W)$ and the renormalized coefficients $C_{i}(\mu)$ can be
seen in ref.~\cite{ALI96}.
The perturbative QCD corrections to the decay rate $\GGAMAXS$ have two 
distinct contributions:
\begin{itemize}
\item Corrections to the Wilson coefficients
$C_i(\mu)$, calculated with the help of the 
 renormalization group equation, whose solution requires the
knowledge of the anomalous dimension matrix in a given order in $\as$.
\item Corrections to the matrix elements of the operators
${\cal O}_i$ entering through the effective Hamiltonian
at the scale $\mu=O(m_b)$.
\end{itemize}
 The leading logarithmic (LL) corrections 
have been calculated 
by several independent groups in the complete operator basis  
 \cite{Ciuchini}. The LL QCD-enhancement in the
decay rates for $\BGAMAXS$ is a good factor, about $2.5$.  
First calculations of the NLO corrections to the anomalous dimension matrix
have been recently reported by Chetyrkin, Misiak and M\"unz 
\cite{Misiak96}. The total influence of the next-to-leading (NLO) anomalous 
dimension on the effective coefficient 
for the $\BGAMAXS$ decay, $C_{7}^{eff}(\mu=m_b)$, is found to be
modest and  not to exceed $6\%$ in decay rate. This  is based
on using the existing results on the NLO matching conditions, i.e.,the 
Wilson coefficients $C_{i}(\mu=m_W)$. Of these, the first six 
corresponding to the four-quark operators have been derived by Buras et 
al.~\cite{Buraswc}, and the remaining two $C_7(\mu=m_W)$ and 
$C_8(\mu=m_W)$ have been worked out by Adel and Yao \cite{Yao94}. These
latter are being recalculated, together with the renormalization scheme
dependence \cite{Greub97}. In what follows we shall use the
NLO anomalous dimension correction from \cite{Misiak96}.    

  The NLO corrections to the matrix elements are now available completely. 
They are of two kinds: 
 \begin{itemize}
 \item QCD Bremsstrahlung corrections $b \to s \gamma + g$, which are
needed both to cancel the infrared divergences
 in the decay rate for
$\BGAMAXS$ and in obtaining a non-trivial QCD contribution to the
photon energy spectrum in the inclusive decay $\BGAMAXS$.
\item Next-to-leading order virtual corrections to the matrix elements
in the decay $b \to s +\gamma$. 
\end{itemize}
The Bremsstrahlung corrections were calculated in \cite{ag1,ag2}  
in the truncated basis and subsequently in the complete 
operator basis \cite{ag95,Pott95}.
The NLO virtual corrections have also been calculated in the meanwhile 
\cite{GHW96}. They have played a key role in reducing the 
scale-dependence of the LL inclusive decay 
width. All of these pieces can be combined to get the NLO decay width
$\GGAMAXS$ and the details can be seen in \cite{Misiak96}.
It is customary to express the
 branching ratio  $\BBGAMAXS$ in terms of the
semileptonic decay branching ratio ${\cal B} (B \to X\ell \nu_\ell)$,
\begin{equation}
\label{brdef}
{\cal B} ( B \ra  X_{s} \g) = [\frac{\Gamma(B \ra  
\gamma + X_{s})}{\Gamma_{SL}}]^{th}
\, {\cal B} (B \to X\ell \nu_\ell), 
\end{equation}  
where the leading-order QCD corrected expression for $\Gamma_{SL}$
 can be seen in \cite{ALI96}.

In addition to the perturbative QCD improvements discussed above, also the
leading power corrections in $1/\mb$ have been calculated to the 
numerator and denominator in eq.~\ref{brdef} \cite{georgi,manoharwise}.
 The leading order $(1/m_b)$ power 
corrections in the heavy quark expansion are identical 
in the inclusive decay rates for  $\BGAMAXS$ and $B \to X \ell \nu_\ell$,
as far as the kinetic energy term is concerned (in the HQET jargon the 
term proportional to $\lambda_1$). The correction proportional to the
magnetic moment term (i.e., involving $\lambda_2$) are not identical in 
the two decay rates but 
differ only marginally. Thus, including or neglecting these corrections
in the above equation makes a difference of only $1\%$ in $\BBGAMAXS$.

Using $|V_{ts}^* V_{tb}/V_{cb}|=0.976 \pm 0.010$ obtained from the unitarity 
constraints and current values of the various input parameters,
the short-distance contribution is estimated in \cite{Misiak96} as:
\begin{equation}\label{smbsgbr}
{\cal B} (\BGAMAXS )= (3.28 \pm 0.33) \times 10^{-4}~,
\end{equation}
where all the errors have been combined in quadrature.
This is compatible with though marginally larger than the present measurement
${\cal B} (\BGAMAXS )= (2.32 \pm 0.67) \times 10^{-4}$ \cite{CLEOrare2}.
This comparison can also be expressed in terms of the relevant  
CKM matrix element ratio, yielding 
\begin{equation}\label{vtscb}
|\frac{V_{ts}^* V_{tb}}{V_{cb}}| = 0.82 \pm 0.11 (\mbox{expt}) \pm 0.04 
(\mbox{th}), \end{equation}
which is within errors consistent with the estimate obtained from 
the unitarity of the CKM matrix, quoted above. Using the present value
of $V_{tb}\simeq 1$ and $\absvcb = 0.0393 \pm 0.0028$ \cite{Gibbons96}, this 
gives \begin{equation}
\absvts = 0.032 \pm 0.006~.
\end{equation}

With the perturbative corrections under control, the question is how 
important are the non-perturbative corrections in $\BGAMAXS$? In a recent
paper, it has been argued by Voloshin \cite{Voloshinbsg} that the 
inclusive decay rate for $\BGAMAXS$ contains a non-perturbative correction,
whose scale is set by $\mc^{-1}$ rather than $\mb^{-1}$. Such terms are
generated in the lowest non-trivial order by the intermediate $c\bar{c}$  
 state with the radiation of a photon and gluon, if
one expands the amplitude in the small gluon-momentum limit. To leading 
order in $\as$ and $O(\Lambda_{\mbox{QCD}}^2/\mc^2)$, Voloshin finds
a correction in the inclusive radiative decay rate \cite{Voloshinbsg}:
\bea
\label{Voloshinest}
\frac{\delta \Gamma(\BGAMAXS)}{\Gamma(\BGAMAXS)} =\frac{1}{9 C_7} 
\frac{\lambda_2}{\mc^2} \simeq -0.025~,
\eea
where $\lambda_2 =1/4(M_{B^*}^2 - M_B^2) \simeq 0.12$ GeV$^2$ 
characterizes the chromomagnetic interaction of the $b$ quark inside the
$B$ hadron. While, numerically, this correction is innocuous, 
one should be a little wary about the above estimate as it is based on using the leading
order power correction.

 Discussing this point further, we note that in the perturbative 
calculations of 
the diagrams discussed by Voloshin, first carried out for the photon
energy spectrum in $\BGAMAXS$ in \cite{ag1,ag2}, one does not encounter 
the term shown in eq.~(\ref{Voloshinest}) in the inclusive decay 
width. The real and virtual gluon bremsstrahlung corrections 
integrating over the appropriate phase space yield 
an inclusive decay width which is finite in the limit $\mc \to 0$ 
(equivalently $m_u \to 0$). One suspects, though it remains to be proven,  
that an appropriately resummed expansion of the power corrections should yield
a smooth behaviour as well, i.e. the resulting expression
for the inclusive decay width is not  
divergent as $\mc \to 0$ or $m_u \to 0$.

 In fact, were contributions like the one in 
eq.~(\ref{Voloshinest}) present, the problem of power corrections 
would become  much more acute in calculating the decay width for the
CKM-suppressed decay $\BGAMAXD$, where the intermediate $u \bar{u}$ state is
not suppressed by the CKM factor, unlike in the decay $\BGAMAXS$. In this
case, expanding in the small gluon momentum limit, one gets formally a correction of
$O(\Lambda_{\mbox{QCD}}^2/m_u^2)$ in the inclusive decay width, which
would completely overwhelm the corresponding perturbative contribution. The very
appearance of the light quark masses in the denominator is a signal that the
effect being calculated is non-perturbative.
 
Our tentative conclusion is that while non-perturbative contributions 
from the  intermediate $u\bar{u}$ 
and $c\bar{c}$ states are certainly present
in the decay widths for $\BGAMAXS$ and $\BGAMAXD$ - the so-called long
distance (LD) contributions -  a straight forward use of the
operator product expansion is not expected to yield
trust worthy results. One is better off by using data to calculate
the LD-amplitudes without invoking any expansion in the
quark masses. 
Alternative estimates of the LD-effects  based on
phenomenological models along these lines have been given in a number of papers. 
They yield typically \cite{DHT95,GP95,Ricciardi}:
\begin{equation} 
\frac{{\cal M}(\BGAMAXS)_{\mbox{LD}}}{{\cal M}(\BGAMAXS)_{\mbox{SD}}}
\leq 0.1 ~.
\end{equation}
 With the perturbative QCD improvement at hand, theoretical 
accuracy of the decay rate for $\BGAMAXS$
is now limited by the non-perturbative LD contributions. They deserve
further study.

The exclusive-to-inclusive ratio $R_{K^*}$ has been worked out in a
number of models. The present estimates are  based on calculating 
the matrix elements of the electromagnetic penguin operator, i.e., they
implicitly assume the SD-dominance, neglecting the LD contribution.
Taken on their face value, different models give a rather 
large theoretical dispersion on $R_{K^*}$. However, one should stress that 
QCD sum rules and models based on quark-hadron duality give values which 
are in good agreement with the CLEO measurements. Some representative 
results are:
 \bea
R_{K^*} &=& 0.20 \pm 0.06 ~~[\mbox{Ball}~\protect\cite{bksnsr}],
\nonumber\\
R_{K^*} &=& 0.17 \pm 0.05 ~~[\mbox{Colangelo et al.}
~\protect\cite{bksnsr}], \nonumber\\
R_{K^*} &=& 0.16 \pm 0.05 ~~[\mbox{Ali, Braun \& Simma}
~\protect\cite{abs93}], \nonumber\\
R_{K^*} &=& 0.16 \pm 0.05 ~~[\mbox{Narison}
~\protect\cite{bksnsr}], \nonumber\\
R_{K^*} &=& 0.13 \pm 0.03 ~~[\mbox{Ali \& Greub}
~\protect\cite{ag1}].
\eea
Summarizing this section, it is fair to conclude that SM gives a good 
account of data in electromagnetic penguin decays
in inclusive rates, photon energy spectrum, and the ratio 
$R_{K^*}$, but non-perturbative contributions deserve further study. 
\section{Inclusive radiative decay \bgamaxd\ and constraints on the CKM 
parameters} %
\par
The quantity of interest in the
decays $B \to X_d + \gamma$ is the high energy part of the photon energy 
spectrum, which has to be measured requiring that
 the hadronic system $X_d$ recoiling against the
photon does not contain strange hadrons to suppress the large-$E_\g$
photons from the decay $\BGAMAXS$. Assuming that this is feasible,
one can determine  from the ratio of the decay rates
$\BBGAMAXD/\BBGAMAXS$ the parameters of the CKM
matrix. This measurement was first proposed in
\cite{ag2}, where the photon energy spectra were also worked out.

\indent
 In close analogy
with the \bgamaxs\ case discussed earlier,
the complete set of dimension-6 operators relevant for
the processes $b \to d \gamma$ and $b \to d \gamma g$ 
can be written as:
\begin{equation}
\label{heffd}
{\cal H}_{eff}(b \to d)=
 - \frac{4 G_{F}}{\sqrt{2}} \, \xi_{t} \, \sum_{j=1}^{8}
C_{j}(\mu) \, \hat{O}_{j}(\mu),\quad
\end{equation}
where $\xi_{j} = V_{jb} \, V_{jd}^{*}$ with $j=t,c,u$. The operators
 $\hat{O}_j, ~j=1,2$, have implicit in them CKM factors.
We shall use the Wolfenstein parametrization \cite{Wolfenstein},   
in which case the matrix is determined in terms of the four parameters
$A, \lambda=\sin \theta_C$, $\rho$ and $\eta$, and one can express the above
factors as :
\begin{equation} 
\xi_u = A \, \lambda^3 \, (\rho - i \eta),
~~~\xi_c = - A \, \lambda^3 ,
~~~\xi_t=-\xi_u - \xi_c.
\end{equation}
We note that all three CKM-angle-dependent quantities
$\xi_j$ are of the
same order of magnitude, $O(\lambda^3)$. It is convenient to define 
the operators $\hat{O}_1$ and 
$\hat{O}_2$ entering in ${\cal H}_{eff}(b \to d)$ as follows \cite{ag2}:
\begin{eqnarray}
\label{basis}
&&\hat{O}_{1} =
 -\frac{\xi_c}{\xi_t}(\bar{c}_{L \beta} \go{\mu} b_{L \alpha})
(\bar{d}_{L \alpha} \gu{\mu} c_{L \beta})\nonumber\\
& & \mbox{} -\frac{\xi_u}{\xi_t}(\bar{u}_{L \beta} \go{\mu} b_{L \alpha})
(\bar{d}_{L \alpha} \gu{\mu} u_{L \beta}) ,\nonumber \\
&& \hat{O}_{2} =
-\frac{\xi_c}{\xi_t}(\bar{c}_{L \alpha} \go{\mu} b_{L \alpha})
(\bar{d}_{L \beta} \gu{\mu} c_{L \beta})\nonumber\\ 
& & \mbox{} -\frac{\xi_u}{\xi_t}(\bar{u}_{L \alpha} \go{\mu}
b_{L \alpha}) (\bar{d}_{L \beta} \gu{\mu} u_{L \beta}) ,
\end{eqnarray}
with the rest of the operators $(\hat{O}_j;~j=3...8)$ 
defined like their
counterparts ${O}_j$ in ${\cal H}_{eff}(b \to s)$, with the obvious 
replacement
$s \to d$. With this choice, the matching conditions $C_j(m_W)$
 and the solutions
of the RG equations yielding $C_j(\mu)$ become
identical for the two operator bases $O_j$ and $\hat{O}_j$.
The essential difference between  $\GGAMAXS$ and $\GGAMAXD$ 
lies in the matrix elements of the first two operators $O_1$ and $O_2$
(in ${\cal H}_{eff}(b \to s)$) and $\hat{O}_1$ and $\hat{O}_2$ (in 
${\cal H}_{eff}(b \to d)$).
The branching ratio  $\BBGAMAXD$ in the SM (including both the SD- and 
LD-contributions) can be written as:
\bea
\label{branstruc}
\lefteqn{\BBGAMAXD = D_1 \lambda^2}
\nonumber\\&&{}
 \{(1-\rho)^2 + \eta^2 -(1-\rho) D_2 - \eta D_3 +D_4  \} , \quad
\eea
where the functions $D_i$ depend on various parameters such 
as $\mt,\mb,\mc,\mu$, and $\as$.
These functions were calculated in the LL approximation in \cite{ag2}  
and since then their estimates have been improved in 
\cite{aag96}, making use of the NLO calculations discussed earlier. We shall
assume, based on the calculations in \cite{DHT95,GP95,Ricciardi}, that the
LD contributions are small.
Once the decay $\BGAMAXD$ has been measured, the branching ratio can be
solved in terms of the CKM-Wolfenstein parameters $\rho$ and $\eta$.

\begin{figure}[htb]
\vskip -1.0truein
\centerline{\epsfxsize 3.0 truein
{\epsffile{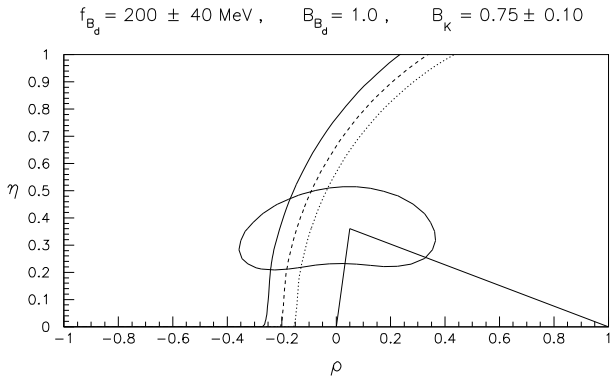}}}
\vskip -1.4truein
\caption[]{Constraints in $\rho$-$\eta$ space from the LEP bound
 $\delms/\delmd > 19.0$. The bounds are presented for 3 choices of
the SU(3)-breaking
parameter: $\xi_s^2 = 1.21$ (dotted line), $1.32$ (dashed line) and $1.44$
(solid line). In all cases, the region to the left of the curve is ruled
out. (Figure taken from \protect\cite{al96}.)}
\label{xslimit}
\end{figure}

 To get an estimate of the inclusive branching 
ratio for $\BGAMAXD$, the CKM parameters $\rho$ and $\eta$ have to be 
constrained from the unitarity fits.
The present experimental and theoretical input
have been summarized  in \cite{al96}.
 The resulting $(\rho$ - $\eta)$ contour is
shown in Fig.~\ref{xslimit} and the values of the input theoretical 
parameters such as the pseudoscalar coupling constant $f_{B_d}$ and the bag
constants $B_K$ and $B_{B_d}$
 are indicated on top of this figure. The fit yields the 
following ranges (at 95\% C.L.) \cite{al96}:
\begin{eqnarray}
 0.20 &\leq & \eta \leq 0.52 , \nonumber \\
 -0.35 &\leq & \rho \leq 0.35 ~.
\label{rhoetarange}
\end{eqnarray}
Including the current lower bound from LEP on
the $B_s^{0}$ - $\overline{B_s^{0}}$ mass difference 
expressed as  
 $\delms/\delmd > 19.0$  \cite{Gibbons96}, one gets further constraints on
$\rho$, with $\eta$ remaining practically unaltered. This is shown 
through the three concentric curves in this figure, which depend on the 
$SU(3)$-breaking parameter   
$\xi_s = f_{B_s} \hat{B}_{B_s}/f_{B_d} \hat{B}_{B_d}$. Using the least 
restrictive of the three values for $\xi_s$ shown in Fig.~\ref{xslimit},
namely $\xi_s =1.1$, the allowed range for $\rho$ is reduced to
$ -0.25 \leq \rho \leq 0.35$.
 %
%
%
%
%
\begin{figure}[htb]
\vskip -1.0truein
\centerline{\epsfxsize 3.0 truein
{\epsffile{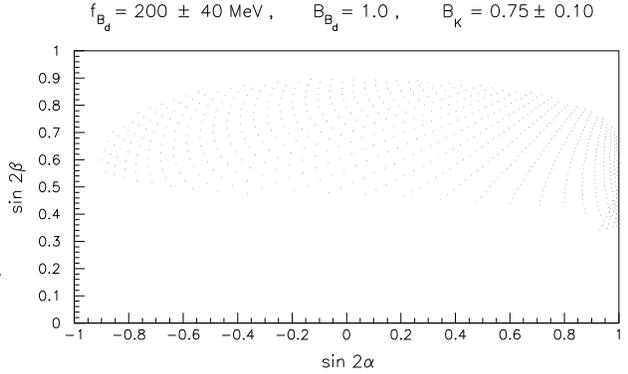}}}
\vskip -1.4truein
\caption[]{Allowed region of the CP-violating quantities $\sin 2 \alpha$ 
and $\sin 2 \beta$ resulting from the CKM fits shown in 
Fig.~\protect\ref{xslimit} including the constraint $\delms/\delmd >19.0$ 
with $\xi_s^2=1.21$. 
 (Figure based on \protect\cite{al96}.)}
\label{albetadelms}
\end{figure}
The preferred CKM-fit values are
$(\rho,\eta) = (0.05,0.36)$ (the unitarity triangle corresponding to these
values is drawn in Fig.~\ref{xslimit}), 
for which one gets \cite{aag96}   \begin{equation}
 \BBGAMAXD = 1.63 \times 10^{-5}.
\end{equation}
Allowing the CKM parameters to vary over the entire allowed domain, one 
gets \begin{equation}
 6.0 \times 10^{-6} \leq \BBGAMAXD \leq 2.8 \times 10^{-5}.
\end{equation} 
The functional dependence of $\BBGAMAXD$ on the CKM parameters
is mathematically different than that of $\delmd/\delms$. However,
qualitatively, the constraints from the measurements of $\delmd/\delms$ and
$\BGAMAXD$ are very similar.  From the experimental point of
view, the situation $\rho <0$ is favourable for both the measurements
of $\BGAMAXD$ and $\delms$.

 We note {\it en passant} that the LEP bound on
the $B_s^{0}$ - $\overline{B_s^{0}}$ mass difference marginally
constrains the otherwise allowed CP-asymmetries in $B$ decays.
 The resulting ($\sin 2\alpha$ - $\sin 2 \beta$) correlation  is shown
in Fig.~\ref{albetadelms}. A comparison with the corresponding correlation 
obtained 
without the LEP $\delms/\delmd$-constraint reveals that the cusp around
$\sin 2 \beta = 0.4$ is now largely gone, but otherwise the allowed ($\sin 2 
\alpha$ - $\sin 2 \beta$) regions are practically the same as in earlier
estimates (see \cite{al96}).
\vspace*{3.0ex}
\section{CKM-suppressed exclusive decays ${\cal B}(B \to V + \gamma )$ } 

\par
Exclusive radiative
 $B$ decays $B \to V + \gamma$, with $V=K^*,\rho,\omega$, are also 
potentially
very interesting from the point of view of determining the CKM parameters
\cite{abs93}. The extraction of these parameters would, however,  involve a 
trustworthy 
estimate of the SD- and LD-contributions in the decay amplitudes.
\par
  The SD-contribution in the 
 exclusive decays $(B^\pm, B^{0}) \to (K^{*\pm}, K^{* 0})+ \gamma$,
$(B^\pm, B^{0}) \to (\rho^\pm,\rho^{0}) + \gamma$,
$B^{0} \to \omega + \gamma$  and the
corresponding $B_s$ decays, $B_s \to \phi + \gamma $, and
$B_s \to K^{* 0} + \gamma $,
involve the magnetic moment operator ${\cal O}_7$ and the related one 
obtained by the obvious change $s \to d$, $\hat{O}_7$.
The transition form factors governing the radiative $B$ decays
 $B \to V + \gamma$ can be generically  defined as:
\be
 \langle V,\lambda |\frac{1}{2} \bar \psi \sigma_{\mu\nu} q^\nu b
 |B\rangle  =
     i \epsilon_{\mu\nu\rho\sigma} e^{(\lambda)}_\nu p^\rho_B p^\sigma_V
F_S^{B\rightarrow V}(0).
\label{defF}
\ee
Here $V$ is a vector meson
with the polarization vector $e^{(\lambda)}$,
$V=\rho, \omega, K^*$ or $\phi$;
$B$ is a generic
$B$-meson $B^\pm, B^{0}$ or $B_s$, and $\psi$ stands for the
field of a light $u,d$ or $s$ quark. The vectors $p_B$, $p_V$ and
$q=p_B-p_V$
correspond to the 4-momenta of the initial $B$-meson and the
outgoing vector
meson and photon, respectively. In (\ref{defF}) the QCD
renormalization of the $\bar \psi \sigma_{\mu\nu} q^\nu b$ operator
is implied.
 Keeping only the SD-contribution 
 leads to obvious relations among the exclusive 
decay rates, exemplified here by the decay
rates for $(B^\pm,B^0) \to \rho + \gamma$ and $(B^\pm,B^0) \to K^* + 
\gamma$: \bea
\lefteqn{\frac{\Gamma ((B^\pm,B^{0}) \to (\rho^\pm,\rho^{0}) + \gamma)}
     {\Gamma ((B_\pm,B^{0}) \to (K^{*\pm},K^{* 0}) + \gamma)}=} 
  \nonumber\\&&{} 
 \frac{\vert \xi_t \vert^2}{\vert\lambda_t \vert ^2}
      \frac{\vert F_S^{B \to \rho }(0)\vert^2}
          {\vert F_S^{B \to K^* }(0)\vert^2} \Phi_{u,d}
  \simeq \kappa_{u,d}\left[\frac{\Vtdabs}{\Vtsabs}\right]^2 \,,
\label{SMKR}
\eea
where $\Phi_{u,d}$ is a phase-space factor which in all cases is close to 1
and
 $\kappa_{i} \equiv [F_S(B_i \to \rho \gamma)/F_S(B_i \to K^* 
\gamma)]^2$.

The LD-amplitudes in radiative $B$ decays from the light quark 
intermediate states necessarily involve other CKM matrix elements. 
Hence, 
the simple factorization of the decay rates in terms of the CKM factors
involving $\Vtdabs$ and $\Vtsabs$ no longer holds thereby
invalidating the relation (\ref{SMKR}) given above. 
In the decays $B \to V + \gamma$ they  are
induced by the matrix elements of the
four-Fermion operators $\hat{O}_1$ and $\hat{O}_2$ (likewise $O_1$ and 
$O_2$). Estimates of these contributions
require non-perturbative methods, which have been studied
in \cite{wyler95,ab95} in conjunction with
 the light-cone QCD sum rule approach to calculate both the SD and LD
--- parity conserving and parity violating --- amplitudes
in the decays $(B^\pm, B^{0}) \to (\rho^\pm,\rho/\omega) + \gamma$.
To illustrate this, we concentrate on the $B^\pm$ decays,
$B^\pm \to \rho^\pm + \gamma$ and take up the neutral $B$ decays
$B^{0} \to \rho (\omega) + \gamma$ at the end.

The LD-amplitude of the four-Fermion operators $\hat{O}_1$, $\hat{O}_2$
is dominated by  the
 contribution of the weak annihilation
of valence quarks in the $B$ meson and it is color-allowed for the
decays of charged $B^\pm$ mesons.
Using factorization, the LD-amplitude in the decay $B_u \to \rho^\pm + 
\gamma$ can be written in terms of the form factors $F_1^L$ and $F_2^L$,
\begin{eqnarray}\label{Along}
{\cal A}_{long} &=&
-\frac{e\,G_F}{\sqrt{2}} V_{ub}V_{ud}^\ast
\left( C_2+\frac{1}{N_c}C_1\right) m_\rho
\varepsilon^{(\gamma)}_\mu \varepsilon^{(\rho)}_\nu
\nonumber\\&&{}\times
 \Big\{-i\Big[g^{\mu\nu}(q\cdot p)- p^\mu q^\nu\Big] \cdot 2 F_1^{L}(q^2)
\nonumber\\
& & \mbox{}  +\epsilon^{\mu\nu\alpha\beta} p_\alpha q_\beta
 \cdot 2 F_2^{L}(q^2)\Big\}\,.
\end{eqnarray}
Again, one has to invoke a model to calculate the form factors.
The parity-conserving and parity-violating amplitudes turn out
to be numerically close to each other in the QCD sum rule approach, 
$F_1^L\simeq F^L_2 \equiv F_L$,
hence the ratio of the LD- and the SD- contributions reduces to a number 
\cite{ab95}
 \begin{equation}\label{ratio2p}
{\cal A}_{long}/{\cal A}_{short}=
R_{L/S}^{B^\pm\to\rho^\pm\gamma}
\cdot\frac{V_{ub}V_{ud}^\ast}{V_{tb}V_{td}^\ast} ~.
\end{equation}
Using $C_2=1.10$, $C_1=-0.235$, $C_7^{\mathit{eff}}=-0.306$
from Ref.~\cite{ALI96} (corresponding to the scale $\mu=5$ GeV) gives:
\bea\label{result2}
R_{L/S}^{B^\pm\to\rho^\pm\gamma} & \equiv &
 \frac{4 \pi^2 m_\rho(C_2+C_1/N_c)}{m_b C_7^{\mathit{eff}}}
\cdot\frac{F_L^{B^\pm \to \rho^\pm \gamma}}{F_S^{B^\pm \to \rho^\pm 
\gamma}}\nonumber\\
&=& \mbox{} -0.30\pm 0.07 ~,
\eea
which is not small.
 To get a ball-park estimate of the ratio
${\cal A}_{long}/{\cal A}_{short}$, we take the central value from 
the CKM fits, yielding $\Vubabs/\Vtdabs \simeq 0.33$ \cite{al96},
\bea
|{\cal A}_{long}/{\cal A}_{short}|^{B^\pm\to\rho^\pm\gamma}
&=& |R_{L/S}^{B^\pm\to\rho^\pm\gamma}|
\frac{|V_{ub}V_{ud}|}{|V_{td}V_{tb}|}
\nonumber\\&&{}
 \simeq 10\% ~.
\label{bpmld}
\eea
Thus, the CKM factors suppress the LD-contributions.

The analogous LD-contributions to the neutral $B$ decays
$B^{0}\to\rho\gamma $ and $B^{0}\to\omega\gamma $ are
expected to be much smaller.
 The corresponding form factors for the decays
$B^{0} \to \rho^0(\omega)  \gamma$ are obtained from
the ones for the decay $B^\pm\to\rho^\pm \gamma$ discussed above by the
replacement of the light quark charges
 $e_u\to e_d$, which gives the factor $-1/2$; in addition,
and more importantly, the
LD-contribution to the neutral $B$ decays
is colour-suppressed, which reflects itself
through the replacement of the factor
$a_1$  by $a_2$. This yields for the ratio
\begin{equation}
\frac{R_{L/S}^{B^{0}\to\rho\gamma}}{R_{L/S}^{B^\pm\to\rho^\pm\gamma}}=
\frac{e_d a_2}{e_u a_1} \simeq -0.13 \pm 0.05 ,
\end{equation}
where the numbers are based on using
$a_2/a_1 = 0.27 \pm 0.10$ \cite{BH95}. This would then yield 
$R_{L/S}^{B^{0}\to\rho\gamma} \simeq R_{L/S}^{B^{0}\to\omega\gamma}=0.05$,
which in turn gives
\begin{equation}
 \frac{{\cal A}_{long}^{B^{0}\to\rho\gamma}}{{\cal 
A}_{short}^{B^{0}\to\rho\gamma}}\leq 0.02.
\end{equation}
This, as well as the estimate in eq.~\ref{bpmld}, should be taken only as 
indicative in view of the approximations made in 
\cite{wyler95,ab95}. That the LD-effects remain small
in ${B^{0}\to\rho\gamma}$ has been supported in a recent analysis
based on the soft-scattering of on-shell hadronic decay products
$B^{0} \to \rho^0 \rho^0 \to \rho \gamma$ \cite{DGP96},
though this paper estimates them somewhat higher   
(between $4 -8\%$).

\begin{figure}[htb]
\vskip -1.6truein
\centerline{\epsfxsize=2.7in
{\epsffile{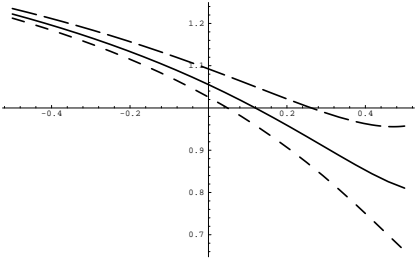}}}
\vskip -1.0truein
\caption[]{
 Ratio of the neutral and charged $B$-decay rates
 $\Gamma (B^\pm \to \rho^\pm \gamma)/2\Gamma (B^{0} \to \rho \gamma)$ as a
function
of the Wolfenstein parameter $\rho$, with $\eta =0.2$ (short-dashed curve),
$\eta =0.3$ (solid curve), and $\eta =0.4$ (long-dashed curve). (Figure taken
from \protect\cite{ab95}.)
\label{abfig2}}
\end{figure}

 The relations, which
obtain ignoring LD-contributions and using isospin invariance,
\beq\label{ratio2}
\Gamma(B^\pm \to \rho^\pm \gamma)=2 ~\Gamma(B^{0}\to \rho^0  \gamma)
    = 2 ~\Gamma (B^{0} \to \omega  \gamma)~,
\eeq
get modified due to the LD-contributions to
\begin{eqnarray}\label{ratio5}
\lefteqn{\frac{\Gamma(B^\pm\to \rho^\pm\gamma)}{2\Gamma(B^{0}\to \rho\gamma)}
=\frac{\Gamma(B^\pm\to \rho^\pm\gamma)}{2\Gamma(B^{0}\to \omega\gamma)}}
\nonumber\\&&{}
 =\left|1+R_{L/S}^{B^\pm\to\rho^\pm\gamma}
\frac{V_{ub}V_{ud}^\ast}{V_{tb}V_{td}^\ast}\right|^2 =
\nonumber\\&&{}
=1+2\cdot R_{L/S} V_{ud}\frac{\rho(1-\rho)-\eta^2}{(1-\rho)^2+\eta^2}
\nonumber\\&&{}
+(R_{L/S})^2 V_{ud}^2\frac{\rho^2+\eta^2}{(1-\rho)^2+\eta^2}\,,
\end{eqnarray}
where $R_{L/S}\equiv R_{L/S}^{B^\pm\to\rho^\pm\gamma}$.  
The ratio
$\Gamma(B^\pm\to \rho^\pm\gamma)/2\Gamma(B^{0}\to \rho\gamma)
(=\Gamma(B^\pm\to \rho^\pm\gamma)/2\Gamma(B^{0}\to \omega\gamma))$
is shown in Fig.~\ref{abfig2}
 as a function of the parameter $\rho$, with
 $\eta= 0.2, ~0.3$  and $0.4$.
This suggests that
a measurement of this ratio would constrain the Wolfenstein parameters
$(\rho, \eta)$, with the dependence on $\rho$ more marked
than on $\eta$. In particular,
a negative value of $\rho$ leads to a
 constructive interference in
$B_u\to\rho\gamma$ decays, while large positive values of $\rho$ give 
a destructive interference.

\par
The ratio of the CKM-suppressed and CKM-allowed
decay rates  for charged $B$ mesons
gets modified due to the LD contributions. Following earlier discussion,
we ignore the LD-contributions in $\Gamma(B \to K^*\gamma)$. The ratio of
the decay rates in question can therefore be written as:
\begin{eqnarray}\label{ratio3}
\lefteqn{\frac{\Gamma(B^\pm\to \rho^\pm\gamma)}{\Gamma(B^\pm\to 
K^{*\pm}\gamma)} = \kappa_u \lambda^2[(1-\rho)^2+\eta^2]
}
\nonumber\\&&{}
\times\Bigg\{
1+2\cdot R_{L/S} V_{ud}\frac{\rho(1-\rho)-\eta^2}{(1-\rho)^2+\eta^2}
\nonumber\\&&{}
+(R_{L/S})^2 V_{ud}^2\frac{\rho^2+\eta^2}{(1-\rho)^2+\eta^2}\Bigg\}\,,
\end{eqnarray}
The dependence of this ratio on $\eta$ is weak 
but more marked on $\rho$.
The effect of the LD-contributions is modest but not negligible, introducing
an uncertainty  
comparable to the $\sim 15\%$ uncertainty in the overall normalization
due to the $SU(3)$-breaking effects in the quantity $\kappa_u$.

%
%

\indent
Neutral $B$-meson radiative decays are less-prone to the LD-effects,
 as argued above, and hence one expects that to a good approximation
(say, better than $10\%$) the ratio of the decay rates for neutral $B$ meson 
obtained in the approximation of SD-dominance remains valid \cite{abs93}:
\begin{equation}
\frac{\Gamma(B^0\to \rho\gamma,\omega\gamma)}{\Gamma(B\to K^*\gamma)}
 = \kappa_d\lambda^2 [(1-\rho)^2+\eta^2]~,
\end{equation}
where this relation holds for each of the two decay modes separately.

 Finally, combining the estimates for the LD- and SD-form factors in
\cite{ab95} and
\cite{abs93}, respectively, and restricting the Wolfenstein
parameters in the range $-0.25 \leq \rho \leq 0.35$ and $ 0.2 \leq \eta
\leq 0.4$, as suggested by the CKM-fits \cite{al96}, we give the
following ranges for the absolute branching ratios:
\begin{eqnarray}\label{ratio4}
{\cal B}(B^\pm\to \rho^\pm\gamma)
&=& (1.5 \pm 1.1) \times 10^{-6} ~,
\nonumber\\{}
{\cal B}(B^{0}\to \rho\gamma) & \simeq & {\cal B}(B^{0}\to \omega \gamma)
\nonumber\\{}
& = & (0.65 \pm 0.35) \times 10^{-6} ~,
\end{eqnarray}
where we have used the experimental value for the branching ratio
${\cal B} (B \to K^* + \gamma)$
\cite{CLEOrare1},
adding the errors in quadrature. The large error reflects the poor
knowledge of the CKM matrix elements and hence experimental determination
of these branching ratios will put rather stringent constraints on the
Wolfenstein parameter $\rho$.

In addition to studying the radiative penguin decays of the $B^\pm$ and
$B^0$ mesons discussed above, hadron machines such as HERA-B will be in a 
position to study the
corresponding decays of the $B_s^0$ meson and $\Lambda_b$ baryon, such as
$B_s^0 \to \phi + \gamma$ and $\Lambda_b \to \Lambda + \gamma$, which have
not been measured so far. We list below the branching ratios in a number of
interesting decay modes calculated in the QCD sum rule approach in 
\cite{abs93}.
\begin{eqnarray}\label{ratio6}
{\cal B}(B_s\to \phi\gamma)
&=&  (4.9 \pm 2.0) \times 10^{-5} ~,
\nonumber\\
\frac{{\cal B}(B_s\to K^*\gamma)}{{\cal B}(B_d\to K^*\gamma)}
 &\simeq & (0.36 \pm 0.14) \left( \frac{\Vtdabs}{\Vtsabs}\right)^2
 \nonumber\\
\Longrightarrow  {\cal B}(B_s\to K^*\gamma) & = & (1.0 \pm 0.6) 
\times 10^{-6} ~,
\end{eqnarray}
where we have used 
${\cal B}(B_s\to \phi\gamma) =  {\cal B}(B_d\to K^* \gamma)$.

The branching ratio
 of the exclusive decay
$\Lambda_b \to \Lambda + \gamma$ has been estimated recently using heavy 
quark symmetry and 
assumptions about the $q^2$-dependence of the form factors \cite{MR97},
yielding
\begin{equation}
{\cal B}(\Lambda_b \to \Lambda + \gamma) = (1.0 - 4.5) \times 10^{-5}\,.
\end{equation}
The estimated branching ratios in a number of inclusive and
exclusive radiative $B$ decay modes are given  in Table \ref{tab1}.

\section{Inclusive rare decays $B \to X_s \ell^+ \ell^-$ in the SM}

\par
The decays \bxsll, with $\ell=e,\mu,\tau$, provide a more sensitive search
strategy for finding new physics in rare $B$ decays
than for example the decay \bxsg , which constrains
 the magnitude of $C_7^{\mathit{eff}}$.
 The sign of $C_7^{\mathit{eff}}$, which
 depends on the underlying physics, is not
determined by the measurement of ${\cal B}(\BGAMAXS)$. This sign, which 
in our convention is negative in the SM, is in general model dependent.
It is known (see for example \cite{AGM94}) that
in SUSY models, both the negative and positive signs are 
allowed as one scans over the allowed SUSY parameter space.
The \bxsll ~amplitude in the standard model
has in addition to the coefficient $C_7^{\mathit{eff}}$ two 
additional terms, arising from the two FCNC four-Fermi operators
 \footnote{This also
holds for a large class of models such as MSSM and the two-Higgs doublet
models but not for all SM-extensions. In LR symmetric models, for example, 
there
are additional FCNC four-Fermi operators involved \cite{LRsymmetry}.}.
Calling their coefficients $C_{9}$ and $C_{10}$, it has been argued in
\cite{AGM94} that the signs and
magnitudes of all three coefficients $C_7^{\mathit{eff}}$, $C_{9}$ and 
$C_{10}$
can, in principle,  be determined from the decays $\BGAMAXS$ and \bxsll .

\par
 The SM-based rates for the decay \bsll , calculated in the free quark decay
approximation, have been known in the LO approximation for some time
\cite{BSGAM}. The LO calculations have the unpleasant
feature that the decay distributions and rates are scheme-dependent.
 The required NLO calculation is in the meanwhile
available, which reduces the scheme-dependence of the LO effects in these
decays \cite{MisiakBM94}. In addition,
long-distance (LD) effects, which are expected to be very important in the
decay \bxsll, have also been estimated from data
 on the assumption that they arise dominantly due to
the charmonium resonances $ J/\psi$ and $\psi'$ through the decay chains
$B \rightarrow X_s J/\psi (\psi',...) \rightarrow X_s \ell^+ \ell^-$.
Likewise, the  leading $(1/{m_b}^2)$ power corrections
to the partonic decay rate and the dilepton invariant mass distribution
have been calculated with the help of the operator product expansion in the 
effective heavy quark theory \cite{falketalbsll}. The results of 
\cite{falketalbsll} have, however, not been confirmed in a recent 
independent calculation
\cite{AHHM96}, which finds that the power corrections in the branching
ratio ${\cal B}(B \to X_s \ell^+ \ell^-)$ are small (typically
$-1.5\%$). The corrections in the dilepton mass spectrum and the FB
asymmetry are also small over a good part of this spectrum. However, the 
end-point dilepton invariant mass spectrum is not calculable in the
heavy quark expansion and will have to be modeled. Non-perturbative 
effects in \bxsll have been estimated using the Fermi motion model in
\cite{Aliqcd}. These effects are found to be small except for the end-point
dilepton mass spectrum where they change the underlying parton model 
distributions significantly and have to be taken into account in
the analysis of data \cite{AHHM96}.
 
The amplitude for \bxsll is calculated in the effective theory
approach, which we have discussed earlier,  by
extending the operator basis of the effective Hamiltonian
introduced in Eq.~(\ref{heffbsg}):
\begin{eqnarray}\label{heffbsll}
& & {\cal H}_{eff}(b \to s + \gamma ; b \to s + \ell^+\ell^- )= \nonumber\\
  && {\cal H}_{eff} (b \to s + \gamma) -\frac{4 G_F}{\sqrt{2}} V_{ts}^* 
V_{tb} \left[ C_9 {\cal O}_9 +C_{10}{\cal O}_{10} \right],
\nonumber\\&&{}
\end{eqnarray}
where the two additional operators are:
\begin{eqnarray}
{\cal O}_9 &=& \frac{\alpha}{4 \pi} \bar{s}_\alpha \gamma^{\mu} P_L b_\alpha 
\bar{\ell} \gamma_{\mu} \ell , \nonumber\\
{\cal O}_{10} &=& \frac{\alpha}{4 \pi} \bar{s}_\alpha \gamma^{\mu} P_L 
b_\alpha \bar{\ell} \gamma_{\mu}\gamma_5 \ell ~.
\end{eqnarray}

The analytic expressions for $C_{9}(m_W)$ and $C_{10}(m_W)$ can be seen
in \cite{MisiakBM94} and will not be given here.
 We recall that the
coefficient $C_9$ in LO is scheme-dependent. However, this is compensated
by an additional scheme-dependent part in the
(one loop) matrix element of ${\cal O}_9$. We call the
sum  $C_9^{\mathit{eff}}$, which is scheme-independent and enters in the 
physical decay amplitude given below,
\begin{eqnarray}
\lefteqn{{\cal M}(b \to s +\ell^+\ell^-) =
 \frac{4 G_F}{\sqrt{2}} V_{ts}^* V_{tb}\frac{\alpha}{\pi}\overline{\cal M},}
 \nonumber\\
\overline{\cal M} &=& C_9^{\mathit{eff}}\bar{s} \gamma^{\mu} P_L b 
\bar{\ell} \gamma_{\mu} \ell
 +C_{10}\bar{s} \gamma^{\mu} P_L b \bar{\ell} \gamma_{\mu}\gamma_5 \ell
\nonumber\\&&{}
- 2C_7^{\mathit{eff}} \bar{s} i\sigma_{\mu \nu} 
\frac{q^\nu}{q^2}(m_bP_R+m_sP_L)b
\bar{\ell} \gamma^{\mu} \ell ,\nonumber\\
&& {}
\end{eqnarray}
with
\begin{equation}
C_9^{\mathit{eff}} (\hat{s}) \equiv C_9\eta({\hat{s}}) + Y(\hat{s}).
\end{equation}
The function $Y(\hat{s})$ is the one-loop matrix element of ${\cal O}_9$
and can be seen in literature \cite{MisiakBM94,ALI96}.
A useful quantity is the  differential FB asymmetry in the c.m.s. of the
dilepton
defined in refs. \cite{amm91}:
\begin{equation}\label{FBasym}
\frac{d {\cal A}(\hat{s})}{d\hat{s}} = \int_0^1 \frac{d{\cal B}}{dz}
                                      -\int_0^{-1} \frac{d{\cal B}}{dz},
\end{equation}
where $z=\cos \theta$, with $\theta$ being the angle between the lepton
$\ell^+$ and the $b$-quark. This can be expressed as:
\begin{eqnarray}
	{{\rm d}{\cal A}(\hat{s}) \over {\rm d}\hat{s}} & = &
	- {\cal B}_{sl} \frac{3 \alpha^2}{4 \pi^2}
        \frac{1}{f(\hat{m}_c)} u^2 (\hat{s})
        \nonumber\\&&{}
C_{10} \left[ \hat{s}{\cal \Re} ( C_9^{\mathit{eff}}(\hat{s})) +
	2 C^{eff}_7 (1 + \hat{m}_s^2) \right] .
	\label{eqn:dasym}
\end{eqnarray}
 The Wilson coefficients
$C^{eff}_7$, $C^{eff}_9$ and $C_{10}$ appearing in the above equation
and the dilepton spectrum 
can be determined from data by solving the partial branching ratio
${\cal B}(\Delta \hat{s})$ and partial FB asymmetry
${\cal A}(\Delta \hat{s})$, where $\Delta \hat{s}$ defines an
interval in the dilepton invariant mass \cite{AGM94}.

 There are
other quantities which one can measure in the decays $B \to X_s \ell^+ 
\ell^-$ to disentangle the underlying dynamics.
 We mention here the longitudinal polarization
of the lepton in \bxsll, in particular in $B \to X_s \tau^+ \tau^-$,
proposed by Hewett \cite{Hewettpol}. In a recent paper, Kr\"uger and Sehgal
\cite{KS196} have stressed that complementary information is contained in
the two orthogonal components of polarization ($P_T$, the component in the
decay plane, and $P_N$, the component normal to the decay plane), both of
which are proportional to $m_\ell/m_b$, and therefore significant
for the $\tau^+ \tau^-$ channel.

\begin{figure}[htb]
\vskip -0.2truein  
\centerline{\psfig{figure=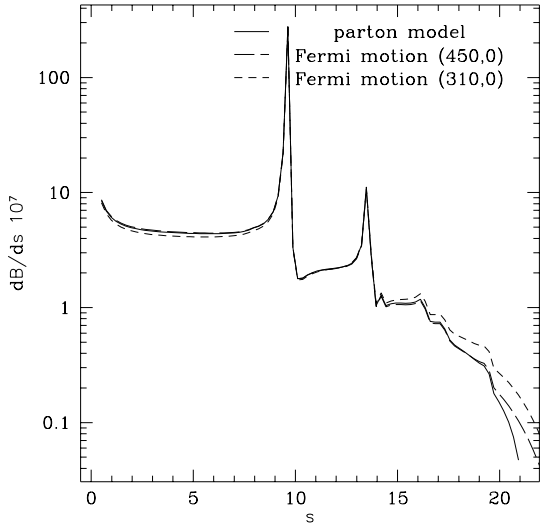,height=8.0cm}}
\vskip -0.1truein
\caption[]{
Dilepton invariant mass distribution in $B \to X_s \ell^+ \ell^-$ in
 the SM including
next-to-leading order QCD correction and LD effects. The solid curve
corresponds to the parton model and the short-dashed and long-dashed
curves correspond to including the Fermi motion effects. The
values of the Fermi motion model are indicated in the figure.
(Figure taken from \protect\cite{AHHM96}).}
\label{fig:dbrnsm}
\end{figure}

Next, we discuss the effects of LD contributions in the
processes $B \to X_s \ell^+ \ell^-$. Note that the
 LD contributions due to the vector mesons such as $J/\psi$ and 
$\psi^\prime$, as well as the continuum $c\bar{c}$ contribution already
discussed, 
appear as an effective $(\bar{s}_L \gamma_\mu b_L)(\bar{\ell} \gamma^\mu 
\ell)$ interaction term only, i.e. in the operator ${\cal O}_9$.
 This implies that the LD-contributions should change
$C_9$ effectively,  $C_7$ as discussed earlier is dominated by the
SD-contribution, and 
$C_{10}$ has no LD-contribution. In accordance with this, 
the function $Y(\hat{s})$ is replaced by,
\begin{equation}
	Y(\hat{s}) \rightarrow Y^\prime(\hat{s}) \equiv Y(\hat{s}) + 
		Y_{\mbox{res}}(\hat{s}),
\end{equation}
where $Y_{\mbox{res}}(\hat{s})$ is given as \cite{amm91},
\bea
	Y_{\mbox{res}}(\hat{s}) &=& \frac{3}{\alpha^2} \kappa 
		\left(3 C_1 + C_2 + 3 C_3 + C_4 + 3 C_5 + C_6 \right)
                \nonumber\\&&{}
	\lefteqn{\times \sum_{V_i = J/\psi, \psi^\prime,...}
		\frac{\pi \Gamma(V_i \rightarrow l^+ l^-) M_{V_i}}{
		M_{V_i}^2 - \hat{s} m_b^2 - i M_{V_i} \Gamma_{V_i}}} ,
\eea
where $\kappa$ is a fudge factor, which appears due to the inadequacy
of the factorization framework in describing data on $B \to J/\psi X_s$.
 The long-distance effects lead to significant interference effects
in the  dilepton invariant mass
distribution  in \bxsll shown in Figs. \ref{fig:dbrnsm}
(likewise in the FB asymmetry). This 
can be used to test the SM, as the signs of the Wilson coefficients in
general are model dependent. For further discussions we refer to 
Ref.~\cite{AHHM96} where also theoretical dispersion on the decay  
distributions due to various input parameters is worked out.
%
%
%
Taking into account the spread in the values of the input parameters,
$\mu, ~\Lambda, ~\mt$, and ${\cal B}_{SL}$
discussed in the previous section in the context of ${\cal B}(B \to X_s + 
\gamma)$, we estimate the following branching ratios for the SD-piece
only (i.e., from the intermediate top quark contribution only) 
\cite{AHHM96}:
 \begin{eqnarray}\label{brbsll}
{\cal B}(\bxsee) &=& (8.4 \pm 2.3) \times 10^{-6}, \nonumber\\
{\cal B}(\bxsmm) &=& (5.7 \pm 1.2) \times 10^{-6}, \nonumber\\
{\cal B}(\bxstt) &=& (2.6 \pm 0.5) \times 10^{-7}, 
\end{eqnarray}
where theoretical errors and the error on ${\cal B}_{SL}$ have been added in 
quadrature.
 The present experimental limit for the inclusive branching ratio in
\bxsll is actually still the one set by the UA1 collaboration some time
ago \cite{UA1R}, namely ${\cal B}(\bxsmm) > 5.0 \times 10^{-5}$. As far
as we know, there are no interesting limits on the other two modes,
involving $X_s e^+e^-$ and $X_s \tau^+ \tau^-$.

\par
 The experimental limits on the decay rates of the exclusive decays $B \to 
(K,K^*) \ell^+ \ell^-$ \cite{BH95,Tomasz95}, while arguably closer to the 
SM-based estimates,
can only be interpreted in specific models of form factors, which hinders
somewhat their transcription in terms of the information on
the underlying Wilson coefficients.
 Using the exclusive-to-inclusive ratios
\begin{equation}
R_{K^{(*)}\ell\ell} \equiv \Gamma(B \to K^{(*)} \ell^+ \ell^-)/\Gamma (B 
\to X_s \ell^+ \ell^-)~,
\end{equation}
 which were estimated in \cite{AGM92} to be
\begin{equation}
R_{K\ell\ell} = 0.07 \pm 0.02~,\nonumber
\end{equation}
\begin{equation}
R_{K^*\ell\ell} =0.27 \pm 0.0.07~,
\end{equation}
the results are presented in Table \ref{tab1}.
 
 In conclusion, the semileptonic FCNC decays $B \to X_s \ell^+
\ell^-$ (and also the exclusive decays)
 will provide very precise tests of the SM, as they will determine
the signs and magnitudes of the three Wilson coefficients, $C_7,
~C_9^{\mathit{eff}}$ and $C_{10}$.
 We note
that both the dilepton mass distribution and the FB asymmetry are sensitive
to non-SM effects. The case of SUSY was first studied in the classic paper
on this subject by Bertolini et al. \cite{Masieroetal}. Since then, more
detailed studies have been reported in
the literature \cite{AGM94,CMW96}. For a recent update of the
SUSY effects in  $B \to X_s \ell^+ \ell^-$, 
we refer to \cite{Gotoetal96} in which it is shown  that the 
distributions in
this decay may be distorted significantly above the SM-related uncertainties,
even if the inclusive decay rates are not significantly
effected. Based on this and earlier studies along the same lines,
it is conceivable that the FCNC decay \bxsll may open a window on new
physics.

\section{Summary and overview of rare $B$ decays in the SM}

\par
 The rare $B$ decay mode $B \to X_s \nu \bar{\nu}$, and some of the
exclusive channels associated with it,
 have comparatively larger branching ratios. The estimated inclusive 
branching ratio in the SM is \cite{AGM92,BBL95,Grossman}:
\begin{equation}\label{bxsnunu}
 {\cal B}(B \to X_s \nu \bar{\nu}) = (4.0 \pm 1.0) \times 10^{-5}~,
\end{equation}
where the main uncertainty in the rates is due
to the top quark mass. The scale-dependence,
which enters indirectly through the top quark mass, has 
been brought under control through the NLL corrections, calculated in
\cite{BuBu93}. The corresponding CKM-suppressed decay $B \to X_d \nu 
\bar{\nu}$ is related by the ratio of the CKM matrix element
squared \cite{AGM92}:
\begin{equation}\label{bxsdnunu}
 \frac{{\cal B}(B \to X_d \nu \bar{\nu})}
  {{\cal B}(B \to X_s \nu \bar{\nu})} = \left[ 
\frac{\Vtdabs}{\Vtsabs}\right]^2 ~.
 \end{equation}
Similar relations hold for the ratios of the exclusive decay rates which
 depend additionally on the ratios of the form factors squared,
which deviate  from unity through $SU(3)$-breaking terms, in close
analogy with the exclusive radiative decays discussed earlier.
 These decays are particularly attractive  probes of
the short-distance physics, as  the long-distance
contributions are practically absent in such decays. Hence, relations
such as the one in (\ref{bxsdnunu}) provide, in principle, one of 
the best methods for the
 determination of the CKM matrix element ratio $\Vtdabs/\Vtsabs$ 
\cite{AGM92}. From the practical point of view, however, these decay 
modes are rather difficult to measure, in particular
at the hadron colliders.
The present
best upper limit stems from ALEPH collaboration \cite{ALEPHwarsaw}:
\begin{equation}\label{bsnunulim}
 {\cal B}(B \to X \nu \bar{\nu}) < 7.7 \times 10^{-4}.
\end{equation}
The estimated branching ratios in a number of inclusive and
exclusive decay modes are given  in Table \ref{tab1}, updating the 
estimates in \cite{ALI96}.

  Further down the entries in Table \ref{tab1} are listed some two-body 
rare decays, such as $(B_s^0, B_d^0) \to \gamma \gamma$, studied in
 \cite{GGTH}, where only the lowest order contributions
are calculated, i.e., without any QCD corrections, and the LD-effects,
which could contribute significantly, are neglected.
The decays $(B_s^0,B_d^0) \to
\ell^+\ell^-$ have been studied in the 
next-to-leading order QCD in \cite{BuBu93}. Some of them,
in particular, the decays $B_s^0 \to \mu^+ \mu^-$ and perhaps also
the radiative decay $B_s^0 \to \gamma \gamma$, have a fighting chance to be
measured at LHC. The estimated decay rates, which depend on the 
pseudoscalar coupling constant $f_{B_s}$ (for $B_s$-decays) and 
$f_{B_d}$ (for $B_d$-decays), together with the present experimental
bounds are listed in Table \ref{tab1}. Since no QCD corrections have been
included in the rate estimates of $(B_{s}, B_{d}) \to \gamma \gamma$,
the branching ratios are rather uncertain.
 The constraints on beyond-the-SM physics that will
eventually follow from these decays are qualitatively similar to the
ones that (would) follow from the decays $\BGAMAXS$ and $B \to X_s \ell^+ 
\ell^-$, which we have discussed at length earlier.

\begin{table*}[htb]
\setlength{\tabcolsep}{1.5pc}
\newlength{\digitwidth}\settowidth{\digitwidth}{\rm 0}
\catcode`?=\active\def?{\kern\digitwidth}
\caption{Estimates of the branching fractions for some FCNC $B$ decays in the
 SM and comparison with experiments}
\begin{center}
\begin{tabular*}{\textwidth}{@{}l@{\extracolsep{\fill}}rrr}
\hline
                  \multicolumn{1}{r}{Decay Modes} 
                 & \multicolumn{1}{r}{${\cal B}$(SM)} 
                 & \multicolumn{1}{r}{Measurements and
                    90\% C.L. Upper Limits} \\
\hline
 
$ (B^\pm,B^{0}) \to X_{s} \gamma $
& $(3.28 \pm 0.33)  \times 10^{-4}$ 
& $(2.32 \pm 0.67) \times 10^{-4}$~\cite{CLEOrare2}\\
$ (B^\pm,B^{0}) \to K^* \gamma $
& $(4.9 \pm 2.0) \times 10^{-5}$ 
& $(4.2 \pm 1.0) \times 10^{-5}$~\cite{CLEOwarsaw}\\
$ (B^\pm,B^{0}) \to X_{d} \gamma $
& $(1.6 \pm 1.2) \times 10^{-5}$ & --\\
$ B^\pm  \to \rho^\pm + \gamma $
& $(1.5 \pm 1.1)  \times 10^{-6}$ & $ < 1.1 \times
10^{-5}$~\cite{CLEOwarsaw}\\ 
$ B^{0}  \to \rho^0 + \gamma $
& $(0.65 \pm 0.35)  \times 10^{-6}$ & $ < 3.9 \times 10^{-5}$~\cite{CLEOwarsaw}\\ 
$ B^{0}  \to \omega + \gamma $
& $(0.65 \pm 0.35)  \times 10^{-6}$ & $ <1.3 \times 10^{-5}$~\cite{CLEOwarsaw}\\
$ B_{s}\to \phi + \gamma $        
& $(4.9 \pm 2.0)  \times 10^{-5}$ &$ <2.9 \times 
10^{-4}$~\cite{ALEPHICHEP94}\\
$ B_{s}\to K^* + \gamma $        
                    & $(1.0 \pm 0.6)  \times 10^{-6}$ & --\\
$\Lambda_b \to \Lambda + \gamma$ & $(2.75 \pm 1.75) \times 10^{-5}$ &
$ < 5.6 \times 10^{-4}$~\cite{ALEPHICHEP94}\\
$ (B_{d},B_{u}) \to X_{s} e^+ e^- $
                    & $(8.4 \pm 2.2)  \times 10^{-6}$ & --\\
$ (B_{d},B_{u}) \to X_{d} e^+ e^- $
                    & $(4.9 \pm 2.9) \times 10^{-7}$ & --\\
$ (B_{d},B_{u}) \to X_{s} \mu^+ \mu^- $
& $(5.7 \pm 1.2)  \times 10^{-6}$ & $  < 3.6 \times 10^{-5}$~\cite{D0warsaw}\\ 
$ (B_{d},B_{u}) \to X_{d} \mu^+ \mu^- $
& $(3.3 \pm 1.9)  \times 10^{-7}$ &  --\\
$ (B_{d},B_{u}) \to X_{s} \tau^+ \tau^- $
& $(2.6 \pm 0.5)  \times 10^{-7}$ & --\\
$ (B_{d},B_{u}) \to X_{d} \tau^+ \tau^- $
& $(1.5 \pm 0.8)  \times 10^{-8}$ &  --\\
$ (B_{d},B_{u}) \to K e^+ e^- $
& $(5.9 \pm 2.3)  \times 10^{-7}$ & $ < 1.2 \times 10^{-5}$~\cite{Tomasz95}\\
$ (B_{d},B_{u}) \to K \mu^+ \mu^- $
 & $(4.0 \pm 1.5)  \times 10^{-7}$ & $ < 0.9 \times 10^{-5}$~\cite{Tomasz95}\\
$ (B_{d},B_{u}) \to K \mu^+ \mu^- $
 & $(4.0 \pm 1.5)  \times 10^{-7}$ & $ < 0.9 \times 10^{-5}$~\cite{Tomasz95}\\ 
$ (B_{d},B_{u}) \to K^* e^+ e^- $
                    & $(2.3 \pm 0.9)  \times 10^{-6}$ & $ < 1.6 \times
10^{-5}$~\cite{Tomasz95}\\
$ (B_{d},B_{u}) \to K^* \mu^+ \mu^- $
& $(1.5 \pm 0.6)\times 10^{-6}$ & $ <2.5 \times 10^{-5}$~\cite{CDF}\\
$ (B_{d},B_{u}) \to X_{s} ~\nu \bar{\nu} $
& $(4.0 \pm 1.0)  \times 10^{-5}$ & $< 7.7 \times 10^{-4}$~\cite{ALEPHwarsaw}\\
$ (B_{d},B_{u}) \to X_{d} ~\nu \bar{\nu} $
                    & $(2.3 \pm 1.5)  \times 10^{-6}$ & -- \\
$ (B_{d},B_{u}) \to K ~\nu \bar{\nu} $
                    & $(3.2 \pm 1.6)  \times 10^{-6}$ & -- \\
$ (B_{d},B_{u}) \to K^* ~\nu \bar{\nu} $
                    & $(1.1 \pm 0.55)  \times 10^{-5}$ & -- \\
$ B_{s} \to \gamma \gamma $
    & $(3.0 \pm 1.0)  \times 10^{-7}$ & $ < 1.1 \times 10^{-4}$~\cite{L3}\\ 
$ B_{d} \to \gamma \gamma $
    & $(1.2 \pm 0.8) \times 10^{-8}$ & $< 3.8 \times 10^{-5}$~\cite{L3}\\
$ B_{s} \to \tau^+ \tau^- $
                    & $(7.4 \pm 1.9)  \times 10^{-7}$ & --\\
$ B_{d} \to \tau^+ \tau^- $
                    & $(3.1 \pm 1.9)  \times 10^{-8}$ & --\\
$ B_{s} \to \mu^+ \mu^- $
                    & $(3.5 \pm 1.0)  \times 10^{-9}$ & $<1.8 \times
10^{-6}$~\cite{CDF}\\
$ B_{d} \to \mu^+ \mu^- $
                    & $(1.5 \pm 0.9)  \times 10^{-10}$ & $< 6.1 \times
10^{-7}$~\cite{CDF}\\
$ B_{s} \to e^+ e^- $
                    & $(8.0 \pm 3.5)  \times 10^{-14}$ & --\\
$ B_{d} \to e^+ e^- $
                    & $(3.4 \pm 2.3)  \times 10^{-15}$ & --\\
\hline
\end{tabular*}
\end{center}
\label{tab1}
\end{table*}

\noindent
\section{Acknowledgements}
 I would like to thank Hrachia Asatrian,  
Christoph Greub, Gudrun Hiller, David London, Tak Morozumi, and Paris Sphicas for helpful 
discussions and input. The warm hospitality extended  
by the organisers of the KEK conference is gratefully acknowledged.


\end{document}